\documentclass[english,prb,manuscript,superscriptaddress]{revtex4}
\usepackage[LGR,T1]{fontenc}
\usepackage[latin9]{inputenc}
\usepackage{color}
\usepackage{textcomp}
\usepackage{amssymb}
\usepackage{graphicx}

\makeatletter

\DeclareRobustCommand{\greektext}{%
  \fontencoding{LGR}\selectfont\def\encodingdefault{LGR}}
\DeclareRobustCommand{\textgreek}[1]{\leavevmode{\greektext #1}}
\DeclareFontEncoding{LGR}{}{}
\DeclareTextSymbol{\~}{LGR}{126}
\providecommand{\tabularnewline}{\\}

\@ifundefined{textcolor}{}
{%
 \definecolor{BLACK}{gray}{0}
 \definecolor{WHITE}{gray}{1}
 \definecolor{RED}{rgb}{1,0,0}
 \definecolor{GREEN}{rgb}{0,1,0}
 \definecolor{BLUE}{rgb}{0,0,1}
 \definecolor{CYAN}{cmyk}{1,0,0,0}
 \definecolor{MAGENTA}{cmyk}{0,1,0,0}
 \definecolor{YELLOW}{cmyk}{0,0,1,0}
}

\makeatother

\usepackage{babel}
\begin{document}

\title{Controlling the Dark Exciton Spin Eigenstates by External Magnetic
Field}

\author{L. Gantz}

\affiliation{Faculty of Electrical Engineering, Technion-Israel Institute of Technology,
Haifa 32000 Israel}

\affiliation{The Physics Department and the Solid State Institute, Technion-Israel
Institute of Technology, Haifa 32000 Israel}

\author{E. R. Schmidgall}

\affiliation{The Physics Department and the Solid State Institute, Technion-Israel
Institute of Technology, Haifa 32000 Israel}

\affiliation{Department of Physics, University of Washington, Seattle WA 98195 United States}

\author{I. Schwartz}

\affiliation{The Physics Department and the Solid State Institute, Technion-Israel
Institute of Technology, Haifa 32000 Israel}

\author{Y. Don}

\affiliation{The Physics Department and the Solid State Institute, Technion-Israel
Institute of Technology, Haifa 32000 Israel}

\author{E. Waks}

\affiliation{The Joint Quantum Institute and the Institute for Research in Electronics
and Applied Physics, University of Maryland}

\author{G. Bahir}

\affiliation{Faculty of Electrical Engineering, Technion-Israel Institute of Technology,
Haifa 32000 Israel}

\affiliation{The Solid State Institute, Technion-Israel Institute of Technology,
Haifa 32000 Israel}

\author{D. Gershoni }

\affiliation{The Physics Department and the Solid State Institute, Technion-Israel
Institute of Technology, Haifa 32000 Israel}

\affiliation{The Solid State Institute, Technion-Israel Institute of Technology,
Haifa 32000 Israel}

\begin{abstract}
We study the dark exciton\textquoteright s behavior as a coherent
physical two-level spin system (qubit) using an external magnetic
field in the Faraday configuration. Our studies are based on polarization-sensitive
intensity autocorrelation measurements of the optical transition resulting
from the recombination of a spin-blockaded biexciton state, which
heralds the dark exciton and its spin state. We demonstrate control
over the dark exciton eigenstates without degrading its decoherence
time. Our observations agree well with computational predictions based
on a master equation model. 
\end{abstract}
\maketitle
\global\long\def\ket#1{|#1\rangle}

Reliable quantum two-level systems (TLS) are the building blocks for
quantum information processing (QIP). Solid state quantum bits (qubits) that can 
also be well-controlled are required for
QIP to become a viable technology. An important prerequisite of a
solid state qubit is that it has a long coherence time, in which its
quantum state is not randomized by spurious interactions with its
environment \cite{DiVincenzo1995,DiVincenzo2000}. Semiconductor quantum
dots (QDs) confine charge carriers into a three dimensional nanometer
scale region, thus acting in many ways as isolated 'artificial atoms,\textquoteright{}
whose properties can be engineered. They are also compatible with
modern microelectronics, making them particularly attractive as solid
state qubits. Many efforts have been devoted to prepare, control,
and measure the quantum states of charge carriers in QDs\cite{Imamoglu1999,PressNat2010,GreilichNat2009,Kim2010,PressNat2008}.
One of the more studied TLS in QDs is their fundamental optical excitation,
which results in a QD confined electron-hole (e-h) pair. Since light
interacts very weakly with the electronic spin, the photogenerated
e-h pair has antiparallel spin projections on the incident light direction\cite{Benny2011a}.
Such an e-h pair is called a bright exciton (BE). The coherent properties
of the BE have been extensively studied \cite{Kodriano2012,DeGreveNat2011,Godden2012}.
The main advantages of the BE qubit are its accessibility to coherent
control by light and its neutrality, which results in insensitivity
to vicinal electrostatic fluctuations. The main disadvantage is in
its short radiative lifetime (\ensuremath{\sim} 1 ns).
In contrast, dark excitons (DEs) - formed by e-h pairs with parallel
spin projections, are almost optically inactive.\cite{PoemNat2010,Smolenski2012}
Due to small BE-DE mixing, induced by the QD deviation from symmetry, DEs may
still have some residual optical activity.\cite{Zielifniski2015,SlavarXiv2016} However,
their radiative lifetimes are orders of magnitude longer than that
of the BEs \cite{Schwartz2015}.
DEs, like BEs, are neutral and therefore have a long spin coherence time \cite{Schwartz2015}. 
Recently, it was demonstrated that the DE can be optically initiated in a coherent state by an ultrashort
resonant optical pulse \cite{Schwartz2015a}, and that its quantum
state can be coherently controlled and reset using short optical pulses
\cite{Schwartz2015,Schmidgall2015}, making it thus an attractive matter spin qubit.

In this work, we present further experimental study of the DE as a
coherent TLS under an external magnetic field and demonstrate full
control over its eigenstates. Even at zero magnetic field, due to the short
range e-h exchange interaction \cite{Poem2007}, the DE spin states are not degenerate. The spin eigenstates are
the symmetric ${\left|{S_2}\right\rangle }$ and anti-symmetric ${\left|{A_2}\right\rangle }$ coherent superpositions of the DE
spin up (${\left|{+2}\right\rangle }$) and spin down (${\left|{-2}\right\rangle }$)
states \cite{Bayer2002}. At non-vanishing external magnetic fields, however, when the
Zeeman splitting is larger than the exchange interaction, the eigenstates
become the ${\left|{+2}\right\rangle }$ and ${\left|{-2}\right\rangle }$
spin states. 

Our experimental data is corroborated by a theoretical model which
produces excellent agreement with measured photoluminescence (PL)
intensity correlations under various magnetic fields and optical excitation
intensities. This agreement shows that the externally applied field
controls the DE as a qubit, without reducing its inherently long
coherence time. 

\begin{figure}
\includegraphics[width=0.9\columnwidth]{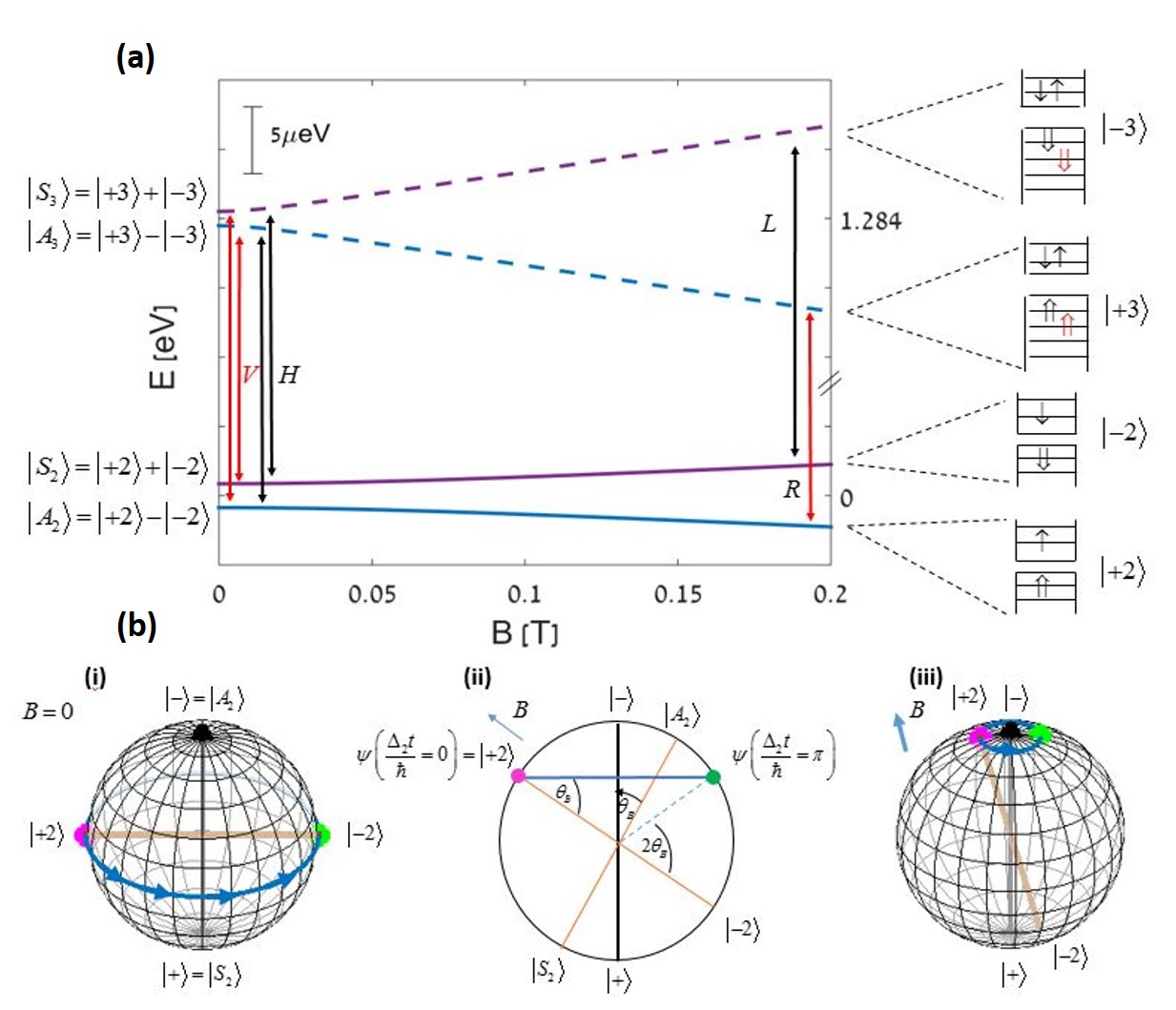}

\protect\caption{\label{fig:1}(a) Schematic description of the energy levels, and
spin wavefunctions of the DE and the $XX_{{T_{3}}}^{0}$ \textendash{}
biexciton as function of an externally applied magnetic field in Faraday
configuration. $\uparrow$($\Downarrow$) represents spin up (down)
electron (hole). The blue and purple solid (dashed) lines represent
the energies of the low and high energy eigenstates of the DE (biexciton)
respectively. The spin eigenstates are written to the right and left
sides of the figure for zero and high field, respectively. Vertical
arrows connecting the DE and biexciton eigenstates mark allowed polarized optical transitions between the eigenstates
at zero and high field. (b) Schematic representation of the changes that
the external field induces on the Bloch sphere of the DE qubit. Shown
are three cases: (i) zero field 
(ii) cross section of the sphere at arbitrary
magnetic field, 
(iii) strong magnetic field. 
The eigenstates  $\left|{A_2}\right\rangle$, $\left|{S_2}\right\rangle$, at zero field and $\left|\pm\right\rangle_2$ at finite field,
and the angle ${\theta_{B}}$ are defined in the text and in EQ. \ref{eq:3}.
The eignestates are always at the poles of the sphere, north pole being the lower energy one. 
The pink dot represents the $\left|{+2}\right\rangle$ state, heralded by detecting R
polarized biexciton photon. The blue circle represents the counter clockwise temporal evolution of the DE state following 
its heralding.}
\end{figure}

At zero magnetic field, due to the short range e-h exchange interaction,
the DE eigenstates are the symmetric $\left|{S_2}\right\rangle =1/\sqrt{2}\left[{\left|{+2}\right\rangle +\left|{-2}\right\rangle }\right]$
and anti-symmetric 
$\left|{A_2}\right\rangle =1/\sqrt{2}\left[{\left|{+2}\right\rangle -\left|{-2}\right\rangle }\right]$
coherent superposition of the spin up (${\left|{+2}\right\rangle }$)
and spin down (${\left|{-2}\right\rangle }$) states, where the anti-symmetric
state is lower in energy \cite{Schwartz2015}. These states are schematically described in Figure.\ref{fig:1} 
The DE can be optically excited, thereby
generating a spin blockaded biexciton $XX_{{T_{3}}}^{0}$ \cite{Kodriano2010}.
This biexciton is comprised of two electrons in a singlet configuration
at their ground level (total spin projection zero), and two holes
with parallel spins forming a triplet (total spin projection \textpm 3),
occupying the ground and second hole levels \cite{Kodriano2010}.
Likewise, as first demonstrated here, the lower and higher eigenstates
of the $XX_{{T_{3}}}^{0}$ qubit are also the anti-symmetric $\left|{A_3}\right\rangle =1/\sqrt{2}\left[{\left|{+3}\right\rangle -\left|{-3}\right\rangle }\right]$
and symmetric
$\left|{S_3}\right\rangle=1/\sqrt{2}\left[{\left|{+3}\right\rangle +\left|{-3}\right\rangle }\right]$
coherent superpositions of the spin up ($\left|{+3}\right\rangle $
) and spin down ($\left|{-3}\right\rangle $ ), respectively. 

The DE and $XX_{{T_{3}}}^{0}$ form an optical \textquotedbl{}$\Pi$-system\textquotedbl{}
since optical transitions are allowed between the ${\left|{+2}\right\rangle }$
(${\left|{-2}\right\rangle }$) DE state to and from the $\left|{+3}\right\rangle $
($\left|{-3}\right\rangle $ ) biexciton state by right (left) handed
circularly polarized light only. At zero magnetic field, the DE and
$XX_{{T_{3}}}^{0}$ eigenstates are therefore optically connected
by linear cross polarized optical transitions denoted as horizontal
(H) and vertical (V), where the H direction is chosen such that it
coincides with the polarization of the ground state BE optical transition
\cite{Kodriano2010}. The system is schematically described in Figure.\ref{fig:1}
(a). 

The time independent Hamiltonian of the DE and the $XX_{{T_{\pm3}}}^{0}$
in the presence of a magnetic field in the Faraday configuration as
expressed in the basis $\left\{ {\left|{+2}\right\rangle ,\left|{-2}\right\rangle ,\left|{+3}\right\rangle ,\left|{-3}\right\rangle }\right\} $
is given by:

{\scriptsize{}
\begin{equation}
\ensuremath{\hat{H}=\frac{1}{2}\left({\begin{array}{cccc}
{-{\mu_{B}}\left({{g_{e}}-{g_{h}}}\right)B} & {\hbar\omega_{2}} & {} & {}\\
{\hbar\omega_{2}} & {{\mu_{B}}\left({{g_{e}}-{g_{h}}}\right)B} & {} & {}\\
{} & {} & {2(\Delta+{\mu_{B}}{g_{2h}}B)} & {\hbar\omega_{3}}\\
{} & {} & {\hbar\omega_{3}} & {2(\Delta-{\mu_{B}}{g_{2h}}B)}
\end{array}}\right)}\label{eq:1}
\end{equation}
}{\scriptsize}
This Hamiltonian represents two decoupled Hamiltonians, one for the
DE and one for the $XX_{{T_{\pm3}}}^{0}$, where ${\mu_{B}}={{e\hbar}\mathord{\left/{\vphantom{{e\hbar}{2{m_{e}}c}}}\right.\kern -\nulldelimiterspace}{2{m_{e}}c}}$
is the Bohr magnetron, B the magnitude of the magnetic field (normal
to the sample surface), $g_{e}$ and $g_{h}$ are the electron and
hole gyromagnetic ratios in the direction of the magnetic field, and
$g_{2h}$ is the gyromagnetic ratio of the two heavy holes in triplet
configuration. 
The sign convention for the gyromagnetic factors is such that positive factors mean that
electron (heavy hole) with spin parallel (antiparallel) to the magnetic
field direction is lower in energy than that with spin antiparallel
(parallel) \cite{Bayer2002}.
We note that the triplet gyromagnetic ratio
is not a simple sum of the gyromagnetic ratios of the individual holes
\cite{Nowak2011}. 
The energy difference between the DE and the $XX_{{T_{\pm3}}}^{0}$ is $\Delta$, and $\hbar{\omega_{2}}$
and $\hbar{\omega_{3}}$ are the energy differences between the DE
and $XX_{{T_{\pm3}}}^{0}$ eigenstates, respectively. All energies
are defined at zero magnetic field. 
From this Hamiltonian, one calculates
the energies and eigenstates of the system. 
Figure. \ref{fig:1}a schematically describes the DE energy level structure,
their magnetic field dependence, and the optical transitions between
their eigenstates. 

The externally applied magnetic field modifies
the eigenstates of both qubits: 
~\cite{Bayer2002}

\begin{equation}
\ensuremath{\begin{array}{l}
{\left|+\right\rangle _{i}}={N^{i}}_{+}\left[{\left|{+i}\right\rangle +\left({\frac{{\beta_{i}}}{{\omega_{i}}}+\sqrt{1+\frac{{\beta_{i}^{2}}}{{\omega_{i}^{2}}}}}\right)\left|{-i}\right\rangle }\right]\\
{\left|-\right\rangle _{i}}={N^{i}}_{-}\left[{\left|{+i}\right\rangle +\left({\frac{{\beta_{i}}}{{\omega_{i}}}-\sqrt{1+\frac{{\beta_{i}^{2}}}{{\omega_{i}^{2}}}}}\right)\left|{-i}\right\rangle }\right]
\end{array}}\label{eq:2}
\end{equation}
where $i=2,3$ $N_{\pm}^{i}$ are normalization factors and
${\beta_{2}}={\mu_{B}}\left({{g_{e}}-{g_{h}}}\right)B$
and ${\beta_{3}}={\mu_{B}}{g_{2h}}B$
are the magnetic energies. 
The energy difference between the two eigenstates
is given by their Zeeman splitting: ${\Delta_{i}}(B)=\sqrt{\beta_{i}^{2}+(\hbar\omega_{i})^{2}}$. If 
one defines $\tan\theta_{B}^{i}=\left({\frac{{\beta_{i}}}{{\hbar\omega_{i}}}}\right)$, 
EQ \ref{eq:2}
can be expressed more conveniently as: 
\begin{equation}
\ensuremath{\begin{array}{l}
\ensuremath{\begin{array}{l}
{\left|+\right\rangle _{i}}=\cos\left({\frac{\pi}{4}+\frac{{\theta_{B}^{i}}}{2}}\right)\left|{+i}\right\rangle +\sin\left({\frac{\pi}{4}+\frac{{\theta_{B}^{i}}}{2}}\right)\left|{-i}\right\rangle \\
{\left|-\right\rangle _{i}}=\cos\left({\frac{\pi}{4}-\frac{{\theta_{B}^{i}}}{2}}\right)\left|{+i}\right\rangle -\sin\left({\frac{\pi}{4}-\frac{{\theta_{B}^{i}}}{2}}\right)\left|{-i}\right\rangle 
\end{array}}\end{array}}\label{eq:3}
\end{equation}
Figure. \ref{fig:1}b presents an intuitive geometrical interpretation for the angle $\theta_{B}$ and the DE Bloch sphere. 
Since in the Faraday configuration the magnetic field direction is aligned with the direction of the $\left|{+2}\right\rangle$ spin state, it follows that $\pi/2-\theta_{B}$  
is the angle between the Bloch sphere eigenstate axis and the direction of the magnetic field. Thus, as the magnitude of the external 
field (B) increases $\theta_{B}^{i}$ approaches  ${\pi\mathord{\left/{\vphantom{\pi2}}\right.\kern -\nulldelimiterspace}2}$ and the eigenstates gradually change their nature. Once the Zeeman
energies significantly exceed the exchange energies, the eigenstates
become the $\left|{\pm2}\right\rangle $ and $\left|{\pm3}\right\rangle $
spin states for the DE and the $XX_{{T_{\pm3}}}^{0}$, respectively.

In self assembled InGaAs QDs, the out of plane g-factors of the electron
and the heavy hole are known to be both positive~\cite{Bayer2002, Alon-Braitbart2006} with that of the electron larger
than that of the hole. 
As a result the lower energy eigenstate contains an increasing contribution from
the $\left|{+2}\right\rangle $ spin state while the higher energy
contains an increasing contribution from the $\left|{-2}\right\rangle $
state, as the magnetic field increases. 
The behavior of the $XX_{{T_{3}}}^{0}$
is similar, because as we show below, the Zeeman splitting of the optical transition from this state to the DE 
is opposite in sign to the Zeeman splitting of the BE transitions.

As can be seen in Figure \ref{fig:1}, at the limit of high magnetic
field, the DE - $XX_{{T_{3}}}^{0}$ system forms two separate TLSs,
in which the DE spin up ($\left|{+2}\right\rangle $) and spin down
($\left|{-2}\right\rangle $) eigenstates are optically connected
to the spin up ($\left|{+3}\right\rangle $) and spin down ($\left|{-3}\right\rangle $)
eigenstates by a right (R) or left (L) hand circularly polarized transition,
respectively. The externally applied field thus changes the polarization
of the optical transitions between the DE and $XX_{{T_{3}}}^{0}$
eigenstates from linearly cross polarized transitions into elliptically
\textendash cross- polarized ones as the field increases and eventually
the optical transitions become cross-circularly polarized.

The state of a TLS (or a qubit) is conventionally described as a point
on the surface of a unit sphere (Bloch sphere). The north pole of
the Bloch sphere describes the lower energy eigenstate and the sphere's
south-pole describes the higher energy eigenstate. The surface of
the sphere describes all possible coherent superpositions of the TLS
eigenstates. Each superposition is therefore uniquely defined by a
polar angle ($\varphi$) and an azimuthal angle ($\theta$):
\begin{equation}
\ensuremath{\left|{\psi}\right\rangle =\cos\left({\frac{\theta}{2}}\right)\left|d\right\rangle +{e^{-i{\varphi^{}}}}\sin\left({\frac{\theta}{2}}\right)\left|u\right\rangle }\label{eq:4}
\end{equation}
where  $\left|d\right\rangle$ ($\left|u\right\rangle$) is the lower (higher) energy eigenstate at the north (south) pole of the Bloch sphere.
When a coherent superposition of a given TLS is formed,
the relative phase between the two eigenstates evolves in time, due
to the energy difference between the two eigenstates $\Delta$ ~\cite{Benny2011a}.
This evolution can be described as a counter-clockwise precession around an axis connecting
the Bloch sphere's poles at a rate $\Delta/\hbar$. The evolution is therefore described such that $\varphi\left(t\right)=\varphi\left({t=0}\right)-{\frac{\Delta}{\hbar}}t$, while $\theta\left(t\right)=\theta\left({t=0}\right)$ remains unchanged
as shown in Figure .1b (i), for the case in which a detection of an R circularly polarized $XX^0_{T_3}$ - biexciton photon initiated the DE in the 
$\left|{+2}\right\rangle$ coherent state. In this case $\theta\left({t=0}\right)=\pi/2$, $\varphi\left({t=0}\right)=0$ and $\Delta/\hbar=\omega_2$.

The externally applied field, induces changes on the DE and $XX_{{T_{3}}}^{0}$
eigenstates as described by EQ. \ref{eq:3}. These changes can be described as geometrical \textquotedbl{}rotations\textquotedbl{}
of their Bloch spheres in space, such that the new direction of the sphere's axis is given by: $\frac{1}{{\Delta_{i}}}\left({{\beta_{i}},0,{\hbar\omega_{i}}}\right)$
where ${\beta_{i}}$, ${\omega_{i}}$ and ${\Delta_{i}}$ are defined
above. Thus, there is an angle ${\theta_B}^i=\tan^{-1}(\beta_{i}/\hbar\omega_{i})$ 
between the sphere's axis in the presence of the external field and the axis in the absence of the field, as described in Figure. 1b (ii).
Relative to the new axis the qubit spin state evolves
in time like
\begin{equation}
\ensuremath{\left|{\psi\left(t\right)}\right\rangle =\cos\left({\frac{\theta}{2}}\right)\left|-\right\rangle_{i} +{e^{-i{\varphi^{}}\left(t\right)}}\sin\left({\frac{\theta}{2}}\right)\left|+\right\rangle_{i} }\label{eq:4}
\end{equation}
Here, detection of an R polarized $XX^{0}_{T_3}$ - biexciton photon, which initiates the DE in the 
$\left|{+2}\right\rangle$ coherent state, defines that $\theta\left({t=0}\right)=\pi/2-\theta_ B$,  $\varphi\left({t=0}\right)=0$ and $\Delta/\hbar=\Delta_2/\hbar$. This situation is schematically described in Figure .1b (ii) and (iii).
%

For probing the DE precession and its dependence on the externally applied magnetic field we used  
continuous wave (CW) resonant optical excitation of the
DE to the  $XX_{{T_{3}}}^{0}$-biexciton.
In the presence of such a CW resonance light field
the two TLSs are
coupled and the time independent Hamiltonian is given by 

{\scriptsize{}
\begin{equation}
\ensuremath{\hat{H}=\left({\begin{array}{cccc}
{-{\Omega_{B}}\left({{g_{e}}-{g_{h}}}\right)} & {{\hbar\omega_{2}}\mathord{\left/{\vphantom{{\omega_{2}}2}}\right.\kern -\nulldelimiterspace}2} & {\hbar\Omega_{R}} & {}\\
{{\hbar\omega_{2}}\mathord{\left/{\vphantom{{\omega_{2}}2}}\right.\kern -\nulldelimiterspace}2} & {{\Omega_{B}}\left({{g_{e}}-{g_{h}}}\right)} & {} & {\hbar\Omega_{L}}\\
{\hbar\Omega_{R}} & {} & {\delta+{\Omega_{B}}{g_{2h}}} & {{\hbar\omega_{3}}\mathord{\left/{\vphantom{{\omega_{3}}2}}\right.\kern -\nulldelimiterspace}2}\\
{} & {\hbar\Omega_{L}} & {{\hbar\omega_{3}}\mathord{\left/{\vphantom{{\omega_{3}}2}}\right.\kern -\nulldelimiterspace}2} & {\delta-{\Omega_{B}}{g_{2h}}}
\end{array}}\right)}\label{eq:5}
\end{equation}
}where 
${{{\Omega_{B}}={\mu_{B}}B}\mathord{\left/{\vphantom{{{\Omega_{B}}={\mu_{B}}B}2}}\right.\kern -\nulldelimiterspace}2}$
, ${\Omega_{R(L)}}$ is the Rabi frequency for right- R (left \textendash{}
L) hand circularly polarized light. 
The detuning of the exciting laser energy from the resonant transition between the DE and $XX^0_{T_3}$ - biexciton is assumed to be zero in our experiments.
EQ. \ref{eq:5} shows that the optical coupling depends on the light
polarization and the spin state of the DE. A circularly polarized
R (L) photon, is absorbed in proportion to the magnitude of the DE
spin state projection on the $\left|{+2}\right\rangle $($\left|{-2}\right\rangle $)
state. 
The $XX_{{T_{\pm3}}}^{0}$-biexciton then starts to precess while it
radiatively recombines into an excited DE state. Detection of a right
(left) hand circularly polarized photon heralds the system
in a well-defined DE state given by $\left|{+2}\right\rangle $($\left|{-2}\right\rangle $).
The DE then precesses until a second photon is
absorbed, and the process repeats itself. 
Therefore, time resolved
intensity autocorrelation measurements of the $XX_{{T_{\pm3}}}^{0}$
spectral line in the circularly polarized basis, provide a straightforward
experimental way for probing the dynamics of the system \cite{Schwartz2015}. 
In the absence of an external field and at low resonant excitation
intensities, such measurements show a temporally oscillating signal
at the frequency $\omega_2$. The visibility of the oscillations
in the degree of circular polarization can be used as a measure for $\theta\left({t=0}\right)$ (where $t=0$
is the time of detecting the first polarized photon), and the phase
of the signal as a measure for $\varphi\left({t=0}\right)$ \cite{Schwartz2015a}.


Our experiments used QDs grown by molecular beam epitaxy (MBE) on
a {[}001{]}-oriented GaAs substrate. One layer of self-assembled InGaAs
QDs was deposited in the center of a one-wavelength microcavity sandwiched
between an upper and lower set of AlAs/GaAs quarter-wavelength layer
Bragg mirrors. The sample was placed inside a tube, immersed in liquid
helium, maintaining a temperature of $4.2$K.
Conducting coil outside the tube was used for generating an external
magnetic field along the tube axis, permitting this way optical studies
in Faraday configuration.
A x60, 0.85 numerical aperture microscope
objective was used to focus the excitation lasers on the sample surface
and to collect the emitted light. We used low intensity high above
bandgap energy 445nm diode CW laser light to photogenerate a steady
state population of DEs in the QD in a statistical manner \cite{Nguyen2012}.
In addition, by using a grating stabilized tunable CW diode laser,
we resonantly excited the DE population in one of
the QDs into a $XX_{{T_{3}}}^{0}$ population
\cite{Schwartz2015}. 

\begin{figure}
\includegraphics[width=1.1\columnwidth]{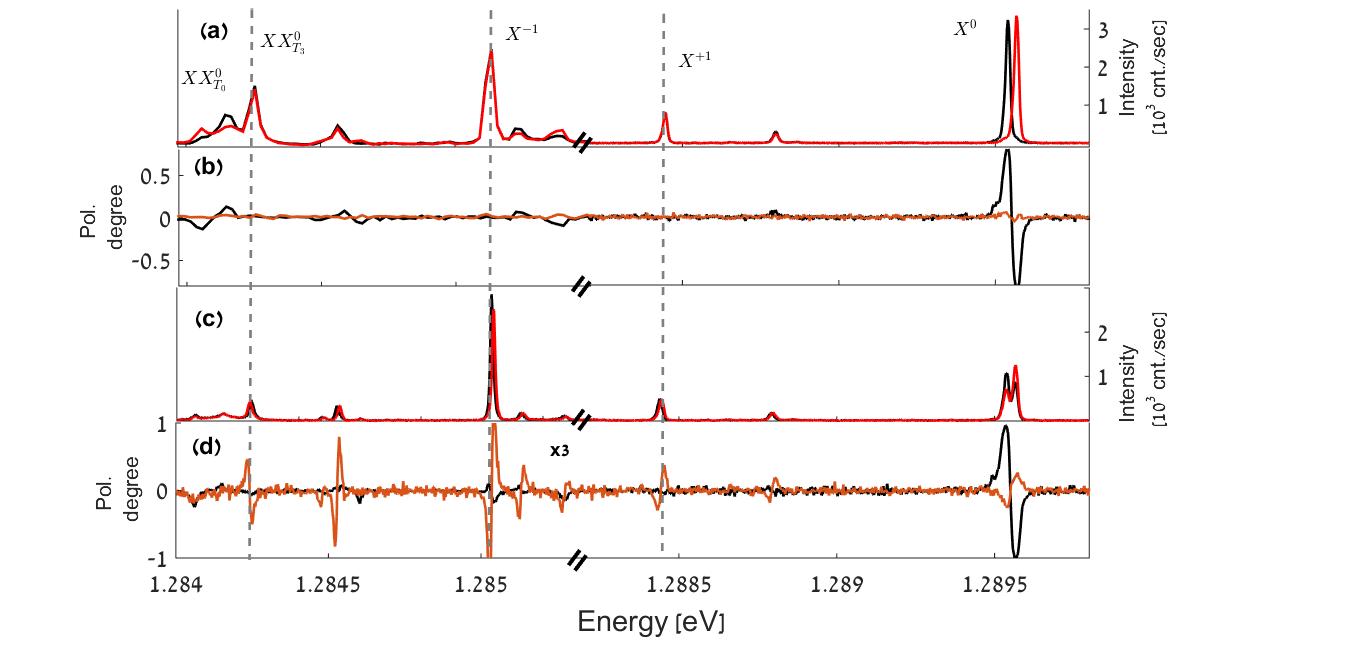}

\protect\caption{\label{fig:2}(a) Rectilinear polarization sensitive PL spectra of
the QD at zero magnetic field. Solid black (red) line represents horizontal
- H (vertical - V) polarization. (b) The degree of rectilinear (black)
and circular (orange) polarizations as a function of the emitted photon
energy. (c) Circular polarization sensitive PL spectra at $B=0.2$T.
Red (black) line represents right-R (left- L) hand circular polarization.
(d) The degree of rectilinear (black) and circular (orange) polarizations
as a function of the emitted photon energy at $B=0.2$T. Note that
the Zeeman splitting of the $XX_{{T_{3}}}^{0}$ line is opposite in
sign to that of the negative, neutral and positive excitons. }
\end{figure}

Figure \ref{fig:2} shows polarization sensitive PL spectra of the
single QD under study. The PL was excited using 445nm non-resonant
cw laser light. Figure \ref{fig:2}a (\ref{fig:2}c) presents the
measured spectra in the two linear (circular) polarizations, in the
absence (presence of $B=0.2$T) external magnetic field. Figure \ref{fig:2}b
(\ref{fig:2}d) presents the obtained degrees of linear (circular)
polarizations as a function of the emitted photon energy in the absence
(presence of $B=0.2$T) external magnetic field. In Figure. 2a the solid
black (red) line represents horizontal - H (vertical - V) polarization
and in Figure \ref{fig:2}c black (red) line represents left-
L (right-R) hand circular polarization. Black (orange) lines in Figure \ref{fig:2}b
and \ref{fig:2}d represent the degree of linear (circular) polarization. The various exciton and biexciton lines are identified in Figure \ref{fig:2}a.

Even in the absence of a magnetic field,
one can clearly observe in Figure \ref{fig:2}a and \ref{fig:2}b that the BE
spectral line is split into two cross linearly polarized components. This split, measured to be $27\pm3 \mu$eV is common to self assembled QDs. 
It results from the anisotropic e-h exchange interaction, mainly due to the QD deviation from cylindrical symmetry.~\cite{Bayer2002,Ivchenko2012} 
The DE degeneracy is also removed mainly due to the short range e-h exchange interaction~\cite{Bayer2002,Ivchenko2012}. However, since the splits $\omega_{2}$ and $\omega_{3}$
are smaller than the radiative linewidth, the linearly polarized
components of the $XX_{{T_{3}}}^{0}$ biexciton line cannot be spectrally resolved. 
Therefore, only one, unpolarized spectral line is observed. 
An upper bound for $\omega_{3}$ < $0.2 ns^{-1} $  corresponding to split of less than $0.82\mu$eV
is deduced directly from the degree of circular polarization memory of the
$XX_{{T_{3}}}^{0}$ biexciton line at zero magnetic field \cite{Schwartz2015}. 
At a sufficiently large magnetic field the line splits into two components.
The lower energy transition is R-circularly polarized and the upper energy one is L-circularly polarized. At a magnitude of $0.2$T,
the splitting amounts to $13.6\pm3$$\mu$eV and it exceeds the 
measured linewidth of $11.4\pm3$$\mu$eV in the absence of external field. 

We note that the measured Zeeman splitting of the $XX_{{T_{\pm3}}}^{0}$
line is opposite in sign to those of the ${X^{+1}}$, the ${X^{-1}}$,
and the ${X^{0}}$ excitonic lines. It follows from simple considerations
that the expected Zeeman splitting of the charged and neutral excitonic
spectral lines is proportional to the sum of the hole and electron
g-factors $\left({{g_{h}}+{g_{e}}}\right)$. Therefore the R polarized part
of these spectral lines is expected to be higher in energy than
the L polarized part. This is indeed what we experimentally observe.
Since the $XX_{{T_{\pm3}}}^{0}$ 
line  splits
in proportion to ${g_{2h}}+{g_{e}}-g_{h}^{*}$ our experimental observations
indicate that the sign of ${g_{2h}}-g_{h}^{*}$ is negative, and its
magnitude in this particular QD is larger than 
that of the electron g-factor. 
These observations are in agreement
with the energy level diagram of Figure. 1a. The dependences of the Zeeman
splitting of the various spectral lines on the g-factors are summarized
in Table \ref{tab:1}.

\begin{table}
\protect\caption{\label{tab:1}The measured Zeeman splitting of various spectral lines. The DE splitting was measured from a similar dot from the same sample. }
\begin{tabular}{|c|c|c|}
\hline 
Line & Zeeman Splitting & Measured at $0.2$T in ($\mu$eV) \tabularnewline
\hline 
\hline 
$X^{0}$ & $\sqrt{(\hbar\omega_{0})^{2}+[\mu_{B}\left(g_{e}+g_{h}\right)B]^{2}}$ & $30\pm3$\tabularnewline
\hline 
$X^{-}/X^{+1}$ & $\mu_{B}(g_{e}+g_{h})B$ & $13.6\pm3$\tabularnewline
\hline 
$XX_{{T_{3}}}^{0}$ & $\mu_{B}\left(g_{2h}+g_{e}-g_{h}^{*}\right)B$ & $13.6\pm3$\tabularnewline
\hline 
$X_{D}^{0}$ & $\mu_{B}(g_{e}-g_{h})B$ & $3.6\pm1$\tabularnewline
\hline 
\end{tabular}
\end{table}

In order to probe the precession of the DE, we excite the sample with
low intensity 445 nm CW laser light. This non-resonant excitation
photogenerates the QD confined BE and DE in a statistical manner.
The BE recombines radiatively within about 1ns, while the DE remains
in the QD until it decays radiatively or an additional charge carrier
enters the QD, whichever comes first. The rate by which additional
carriers enter depends linearly on the power of the (blue) laser
light ($P_{b}$). 
One can tune $P_{b}$ such that the average time between consecutive
arrivals of carriers to the QD is comparable to
the radiative lifetime of the DE. \cite{Schwartz2015}
An additional circularly polarized CW laser light, resonantly tuned
to the DE-$XX_{{T_{\pm3}}}^{0}$ transition is then used for probing the DE precession~\cite{Benny2011}.

\begin{figure}
\includegraphics[width=1\columnwidth]{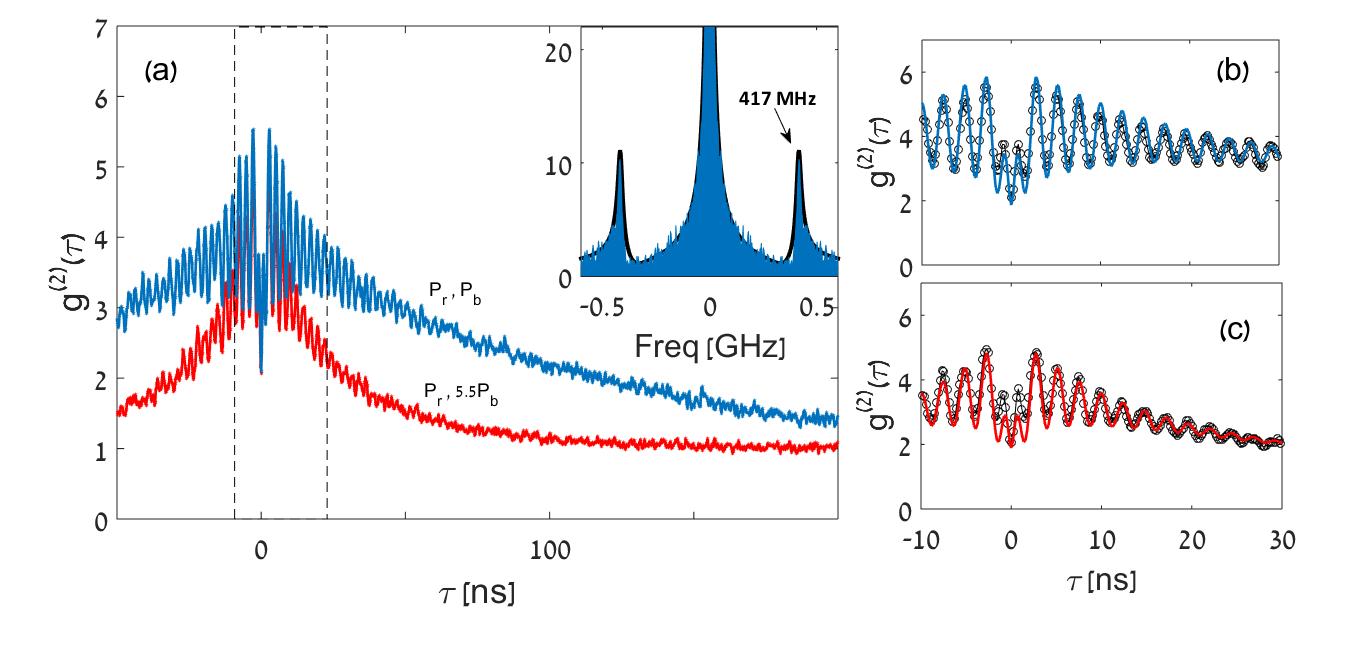}

\protect\caption{\label{fig:3}(a) Intensity autocorrelation measurements ${g^{\left(2\right)}}\left(\tau\right)$
of the emission line $XX_{{T_{3}}}^{0}$without magnetic field. Blue
(red) line represents low- (high-) intensity non-resonant excitation
with blue light $P_{b}=0.1\mu$W ($P_{b}=0.55$\textmu W) and low
intensity resonant excitation with red light ($P_{r}=3.5$\textmu W).
Inset shows the Fourier transform of the low intensity measurement
(blue filled line) and the fitted model calculations (solid black
line), revealing DE precession frequency of $417$MHz \textpm{} $3$MHz
corresponding to a $2.39$ns precession period. (b)-(c) present the
color matched measurements in (a) for a limited temporal window (marked
by dashed vertical lines in (a)). The measured data points (dots)
are overlaid by our model simulations convoluted by the temporal response
of the detectors (solid lines).}
\end{figure}

In Figure \ref{fig:3}(a-c), we present measured and fitted intensity
autocorrelation functions at zero applied magnetic field. As defined,
the functions are normalized to unity at $t\to\infty$. Two measurements 
in two vastly different powers of the blue laser are presented together 
in Figure \ref{fig:3}(a) to clearly demonstrate the reduction of the DE lifetime 
resulting from the increase in the non-resonant excitation power. 
The DE lifetime decreases significantly as the non-resonant blue light
($P_{b}$) power increases as a result of the increase in the flux
of carriers accumulating in the QD. All other experimental conditions,
in particular the intensity of the resonant laser ($P_{r}$), were
kept the same. The measured data points (dots) are overlaid by our
model best fits (continuous lines in Figure \ref{fig:3}(b,c)) using
the parameters listed in Table \ref{tab:2} convoluted with the system
temporal response function \cite{PoemNat2010}. The temporal oscillations in
the correlation function resulting from  the precession of the DE~\cite{Schwartz2015}, 
are clearly observed as well.
The inset to Figure \ref{fig:3}(a) shows the Fourier transform of
the measured and calculated correlation functions under weak blue
excitation. From these measurements we calculate a precession frequency
of $417\pm3$MHz, which corresponds to a precession period of $2.39\pm0.03$
ns and a natural splitting of $1.7\pm0.02$ \textgreek{m}eV between
the two DE eigenstates. 
This splitting, in the absence of magnetic field is due to the short-range e-h exchange interaction \cite{Bayer2002,Zielifniski2015}.
The measured full width at half maximum of the DE frequency at this intensity is $25$ MHz and it 
increases with the excitation power of both the blue and red laser.
This power induced broadening is a consequence of the polarization
oscillation decay, induced by the resonant CW excitation. Much longer
polarization decay times are measured under pulsed excitation \cite{Schwartz2015}.

\begin{figure}
\includegraphics[width=0.5\columnwidth]{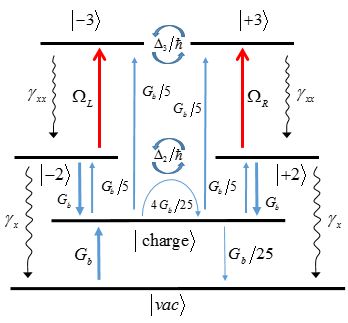}
\protect\caption{\label{fig:4}Schematic description of the levels used in our model.
The DE \textendash{} $XX_{{T_{3}}}^{0}$ biexciton form a $\Pi$-system,
with circularly polarized selection rules for optical transitions. Upwards red arrows represent resonant excitation and curly downwards arrows represent
spontaneous emission. The
charge level represents all states which do not participate in the optical transitions, in particular singly positive or negative charge. 
The non-resonant optical charging and discharging rates, marked by upward and downwards vertical blue arrows 
are proportional to the non resonant excitation rate $G_b$. They were deduced from a set of power dependent measurements 
at zero magnetic field. The
various rates are defined in Table \ref{tab:2}.}
\end{figure}

To model our measurements, we added to the Hamiltonian
presented in Eq \ref{eq:5} a vacuum state and a charge state
as shown in Figure \ref{fig:4}, 
which schematically describes the various states of the system and the transition rates between these states. 
For the sake of simplicity, we only included one additional auxiliary 
charged state in our model. This charge level represents all states which do not participate in the optical transitions, such as a singly positive or negative charged QD. 
With this degree of simplicity, however, we had to estimate the various proportionality constants to $G_{b}$ by which the charged state is
connected to other states by the blue laser excitation (see Figure \ref{fig:4}).
These non-resonant optical charging and discharging rates, marked by upward and downwards vertical blue arrows 
are proportional to the non resonant excitation rate $G_b$. They were deduced from a set of power dependent measurements 
at zero magnetic field. The various rates used in our model are defined in Table \ref{tab:2}.

We then solved the system's master equation, which
includes a Lindblad dissipation part in addition to the Hamiltonian
{\footnotesize{}
\begin{equation}
\ensuremath{\frac{d}{{dt}}\rho\left(t\right)=-\frac{i}{\hbar}\left[{H,\rho\left(t\right)}\right]+\sum\limits _{k}{\left({{L_{k}}\rho\left(t\right)L_{k}^{^{\dag}}-\frac{1}{2}\rho\left(t\right)L_{k}^{^{\dag}}{L_{k}}-\frac{1}{2}L_{k}^{^{\dag}}{L_{k}}\rho\left(t\right)}\right)}}\label{eq:6}
\end{equation}
}where $L_{k}$ represents the various non-Hermitian dissipation rates.
The various parameters used as input to the model are listed and referenced
in Table 2. $G_{b}$ in Figure \ref{fig:4} represents the rate by
which electrons and holes are equally added non coherently to the
QD by the non-resonant blue laser excitation and it is therefore proportional
to the power of  blue laser ($P_{b}$). Since the
DE radiative lifetime is very long, $G_{b}$ essentially defines the
DE lifetime, and the probability to find a DE in the QD. Therefore
$G_{b}$ can be deduced directly from the decay
of the autocorrelation measurements to its steady state (see Figure
\ref{fig:3}). Likewise $\Omega_{R(L)}$ was set proportional to the
square root of the R (L) circularly polarized red laser power
$P_{r}$, as deduced from the power needed to saturate the PL under
excitation with this source. At saturation the co (cross)-circular
intensity autocorrelation signal exhibits no oscillations,
while the lower the power is, the more oscillations are observed. 
This feature facilitated quite sensitive fitting of 
$\Omega_{R(L)}$  so that the observed and calculated number
of oscillations match.

We use the quantum regression theorem \cite{Loudon2000} to solve the master
equation and thus to describe the temporal evolution of the system.
From the numerical solution, we calculated the polarization sensitive
intensity autocorrelation measurement of the  $XX_{T_{\pm3}}^{0}$
line, where detection of the first photon sets the initial system
conditions, and the time by which the second photon is detected defines
the time by which the system evolution is calculated \cite{Regelman2001}.
The calculations were repeated for various blue light and resonance
excitation intensities, and as a function of the magnitude of the
externally applied magnetic field.

\begin{figure}
\includegraphics[width=1\columnwidth]{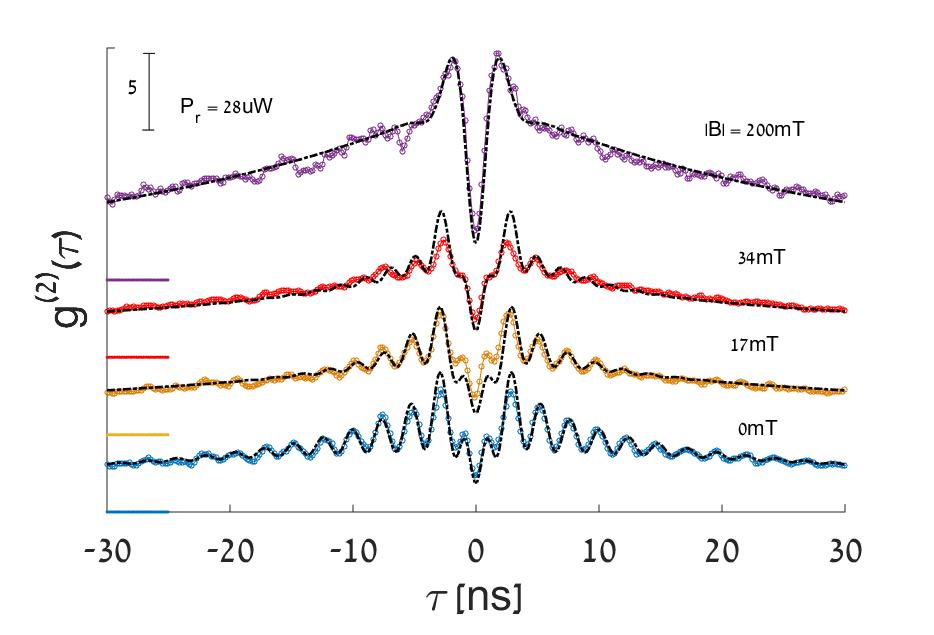}
\protect\caption{\label{fig:5}Measured (symbols) and calculated (lines) circularly
co-polarized intensity autocorrelation functions (${g^{\left(2\right)}}\left(\tau\right)$)
of the emission from the $XX_{{T_{3}}}^{0}$ under weak non-resonant and resonant excitation condition for various externally
applied magnetic fields in Faraday configuration. The solid lines present the best fitted calculations convoluted
with the temporal response of the detectors. The curves are vertically
shifted for clarity and the zero for each measurement is marked by
a color-matched horizontal line. }
\end{figure}

Figure \ref{fig:5} shows co-circular polarization
sensitive intensity autocorrelation measurements of the emission from the $XX_{T_{\pm3}}^{0}$ biexciton
line under weak non-resonant ($P_{b}$) and resonant ($P_{r}$) excitation powers,
for various externally applied magnetic fields. 
Here as well, the measured data points (dots) are overlaid
by our model simulations (dashed lines), convoluted by the detectors temporal response.

\begin{figure}
\includegraphics[width=1\columnwidth]{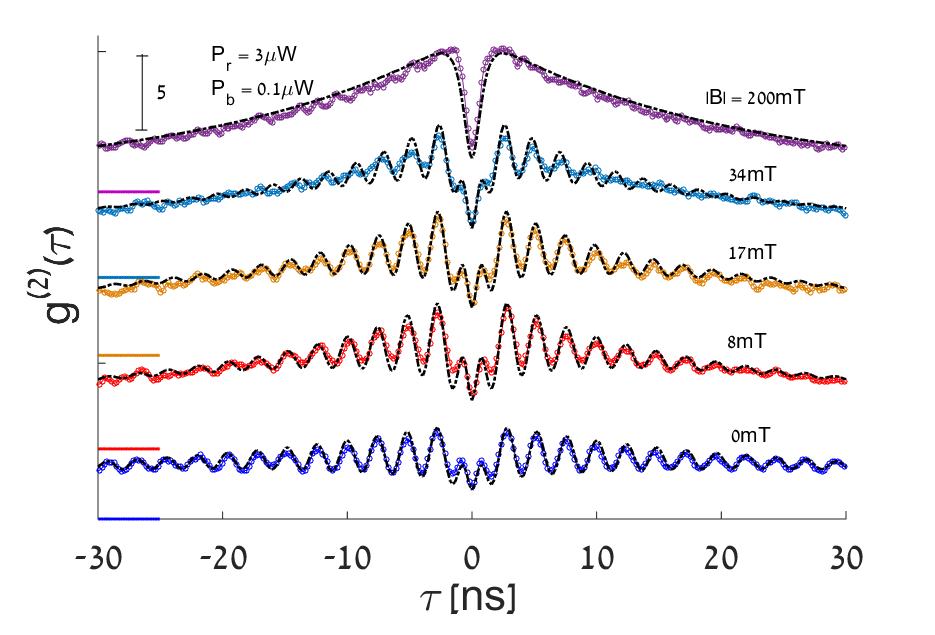}
\protect\caption{\label{fig:6}Measured (symbols) and calculated (lines) circularly
co-polarized intensity autocorrelation functions (${g^{\left(2\right)}}\left(\tau\right)$)
of the emission from the $XX_{{T_{3}}}^{0}$ line under various externally
applied magnetic fields in Faraday configuration, under quasi-resonant
excitation. The solid lines present the best fitted calculations convoluted
with the temporal response of the detectors. The curves are vertically
shifted for clarity and the zero for each measurement is marked by
a color-matched horizontal line. }
\end{figure}

\begin{table}
\protect\caption{\label{tab:2}Physical values used in model calculations. g-factors for the electron and hole were taken from ref [23] and slightly modified to best fit our measurements (see table \ref{tab:1}). }
\begin{tabular}{|c|c|c|c|c|}
\hline 
Parameter & Symbol & Value & Unit & Obtained by\tabularnewline
\hline 
\hline 
$XX_{{T_{3}}}^{0}$ lifetime & ${1\mathord{\left/{\vphantom{1{\gamma_{xx}}}}\right.\kern -\nulldelimiterspace}{\gamma_{xx}}}$ & $0.7$ & nsec & Measured\tabularnewline
\hline 
{DE lifetime} & 
{${1\mathord{\left/{\vphantom{1{\gamma_{x}}}}\right.\kern -\nulldelimiterspace}{\gamma_{x}}}$} & 
{$1000$} & 
{nsec} & 
{Measured}\tabularnewline
\hline 
$XX_{{T_{3}}}^{0}$ precession rate & ${\omega_{3}}$ & $2\pi/6.25$ & 
{rad/nsec} & Measured lower bound\tabularnewline
\hline 
DE precession rate & ${\omega_{2}}$ & $2\pi/2.39$ & 
{rad/nsec} & Measured\tabularnewline
\hline 
Electron g-factor & ${g_{e}}$ & $0.72\pm0.12$ & dimensionless & Measured from different dot. \tabularnewline
\hline 
Hole g-factor & ${g_{h}}$ & $0.41\pm0.15$ & dimensionless & Measured from different dot. \tabularnewline
\hline 
Triplet hole g-factor & ${g_{2h}}$ & $-0.578\pm0.01$ & dimensionless & Measured Zeeman splitting\tabularnewline
\hline 
\end{tabular}

\end{table}

The specific DE-biexciton resonance that we discussed so far is such that an electron is added
to the first level and a heavy hole is added to the second level
thereby directly exciting the $XX_{T_{\pm3}}^{0}$ biexciton.
The use of this resonance is not very convenient for two reasons: a) The oscillator strength of the resonance is relatively weak
due to the different parities of the electron and heavy-hole envelope functions. b) The width of this resonance is relatively narrow, since it is set by the radiative recombination lifetime of the state ($700$ ps).
As a result, excitation to this resonant is very sensitive to the detuning 
from resonance, which becomes
highly sensitive to the externally applied magnetic field.  
We therefore repeated the measurements using a DE-biexciton resonance in which, 
as before, the electron is added to the first level but the heavy hole is added to the fourth level.
This excited biexciton state has significantly larger oscillator strength, 
since the electron and hole envelope wavefunctions are of same parity.
Moreover, this excited biexciton state relaxes non-radiatively,
by a spin conserving process in which a phonon is emitted, to the $XX_{{T_{3}}}^{0}$
ground state. The process occurs within $70$ps (see Supplementary
Information of Ref 12). 
As a result, the width of the resonance is significantly broader than that of the  $XX_{{T_{3}}}^{0}$, and consequently 
its excitation is less sensitive to detuning and to variations in the externally applied field. 
These advantages, make the experiments less demanding, while hardly affecting our conclusions regarding the influence of the 
externally applied field on the DE as a qubit.

Figure \ref{fig:6} shows, co-circular polarization
sensitive intensity autocorrelation measurements of the $XX_{T_{\pm3}}^{0}$
emission line for various magnetic field intensities, under fixed
weak non-resonant ($P_{b}$) and quasi-resonant ($P_{r}$) excitation powers.
The measured data points (dots) are overlaid by our model simulations
(dashed lines).  For these simulations the excited biexciton levels were added to the model, together with their non-radiative, spin preserving
relaxation channels. 
The observed reduction in the visibility of the oscillations as the magnetic field increases 
is observed in both resonant and quasi-resonant excitations. The source of this reduction is 
explained in Figure \ref{fig:1}(b) as resulting from the field induced
changes in the DE qubit eigenstates. 
For example, at a field of $B=0.2$T the DE splitting
was calculated in Figure \ref{fig:1} to be $4$\textmu eV, which is
larger than the measured zero magnetic
field splitting of $1.7$\textmu eV. Hence, as expected, no oscillations are observed, and the system
can be described as two separated TLSs.

\begin{figure}
\includegraphics[width=1\columnwidth]{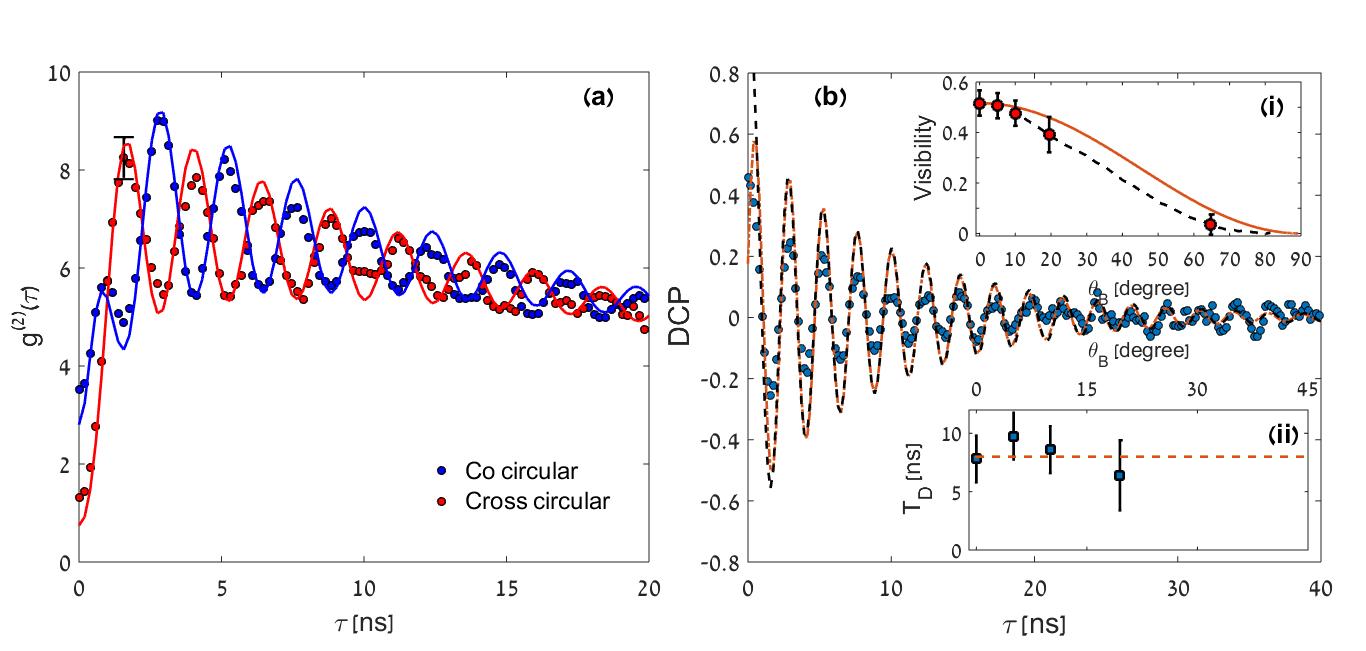}
%
%
\protect\caption{\label{fig:7}(a) Measured (points) and fitted (solid lines) polarization
sensitive correlation functions of the co- (blue) and cross- (red)
circular polarization intensity autocorrelation at $B=8$mT. Solid
lines are the results of our numerical model best fitted calculations
convoluted with the temporal response of the detectors. (b) Measured
(blue dots - obtained from (a)), and calculated (orange line - obtained
from the calculations without the convolution with the detector response),
time resolved DCP. The dashed black line represents
the best fitted analytical expression (Eq. \ref{eq:9})
to the numerical model. The upper inset shows the visibility of the polarization
oscillations as a function of the magnetic field. 
The dashed black line describes the expected dependence as deduced from the numerical calculations. 
The solid orange line describes the analytical expression following Eq \ref{eq:10}. The lower inset shows the
polarization decay time ($T_{D}$) of the DCP as a function of the magnetic field. }
\end{figure}

In Figure \ref{fig:7}(a), we present as an example, the measured (points) and best
fitted model calculations (convoluted with the detector response,
solid lines) polarization sensitive intensity auto correlation functions
of the $XX_{T_{\pm3}}^{0}$ line at $B=8$mT for the quasi-resonant
excitation case. The blue (red) color represents co- (cross-) circular
polarizations of the first and second detected photon. From the two
autocorrelation functions $g_{\parallel}^{\left(2\right)}\left(\tau\right)$
and $g_{\bot}^{\left(2\right)}\left(\tau\right)$, the temporal response
of the degree of circular polarization 
(DCP)
$D\left(\tau\right)$
can be readily obtained:

\begin{equation}
\ensuremath{D\left(\tau\right)=\frac{{g_{\parallel}^{\left(2\right)}\left(\tau\right)-g_{\bot}^{\left(2\right)}\left(\tau\right)}}{{g_{\parallel}^{\left(2\right)}\left(\tau\right)+g_{\bot}^{\left(2\right)}\left(\tau\right)}}
}\label{eq:7}
\end{equation}
In Figure \ref{fig:7}(b), we present $D\left(\tau\right)$,
obtained from Figure \ref{fig:7}(a), where data points present the
measured value and the orange dashed line represents the DCP obtained from the best fitted
numerical model without convolving the detector response function. 

The DCP can be also obtained analytically, using the following considerations:
Recalling that the DCP is given by: 
\begin{equation}
\ensuremath{D\left(\tau\right)=\frac{{{{\left|{\left\langle {+2}\right.\left|{\psi\left(\tau\right)}\right\rangle }\right|}^{2}}-{{\left|{\left\langle {-2}\right.\left|{\psi\left(\tau\right)}\right\rangle }\right|}^{2}}}}{{{{\left|{\left\langle {+2}\right.\left|{\psi\left(\tau\right)}\right\rangle }\right|}^{2}}+{{\left|{\left\langle {-2}\right.\left|{\psi\left(\tau\right)}\right\rangle }\right|}^{2}}}}}
\label{eq:8}
\end{equation}
and substituting $\left|{\psi\left(\tau\right)}\right\rangle $ using Eq \ref{eq:3} and Eq \ref{eq:4} 
one obtains:
\begin{equation}
\ensuremath{D\left(\tau\right)=\left[{{{\cos}^{2}}\left({\frac{{{\Delta_{2}}\tau}}{{2\hbar}}}\right)-{{\sin}^{2}}\left({\frac{{{\Delta_{2}}\tau}}{{2\hbar}}}\right)\cos\left({2{\theta_{B}}}\right)}\right]
}\label{eq:8}
\end{equation}
Eq \ref{eq:8} describes the temporal evolution of the DCP assuming that the first detected photon is R polarized ($\varphi (\tau =0)=0$ in Eq \ref{eq:3} ),
the radiative decay is instantaneous, and the coherence of the DE is infinitely long. 
The fact that the biexciton precesses and has a finite radiative lifetime  ($\tau_R=700 ps$) adds a prefactor $A_v=0.84$ to Eq \ref{eq:8}. This prefactor was deduced directly from the polarization memory measurements. Assuming, in addition, that the DCP decays exponentially with a characteristic time $T_D$, due to the optical re-excitation and the decoherence of the DE, transforms Eq \ref{eq:8} into: 

\begin{equation}
\ensuremath{D\left( \tau  \right) = {A_V}\left[ {{{\cos }^2}\left( {\frac{{{\Delta _2}\tau }}{{2\hbar }}} \right) - {{\sin }^2}\left( {\frac{{{\Delta _2}\tau }}{{2\hbar }}} \right)\cos \left( {2{\theta _B}} \right)} \right] \cdot {e^{ - {\tau  \mathord{\left/
 {\vphantom {\tau  {{T_D}}}} \right.
 \kern-\nulldelimiterspace} {{T_D}}}}}}
\label{eq:9}
\end{equation}

From Eq \ref{eq:9} the visibility of the DCP oscillations and its dependence on the magnetic field can be straightforwardly calculated for the case $\tau<<T_D$:     
\begin{equation}
V(\theta_B)=[D(\theta_B)_{max} - D(\theta_B)_{min}]/2 =A_v^{'}(1+\cos2\theta_B)/2 =A_v^{'}\cos^2\theta_B 
\label{eq:10}
\end{equation}
where $ D(\theta_B)_{max}$ and $ D(\theta_B)_{min}$ are obtained from Eq \ref{eq:9} for $\frac{{{\Delta_{2}}\tau}}{{2\hbar}}=0$ and 
$\frac{{{\Delta_{2}}\tau}}{{2\hbar}}=\pi$, respectively, and $A_v^{'}<A_v$, includes corrections due to the exponential decay of the DCP.  

The best fitted numerical model to the data of each of the measurements, presented in Figure
\ref{fig:6}, represents the measured evolution of the DE after  
the quantified finite temporal response of the experimental setup was considered. Therefore, to the best numerical model fits, those without the convoluted spectral response of the system, 
we fitted the analytical expression of Eq \ref{eq:9}, as shown in Figure \ref{fig:7} (b) by the dashed black line.

The observed decay of the DCP ($T_{D}$) has two main contributions.
The first one results from the actual decoherence of the DE spin qubit
due to its interaction with the nuclei spins $T_{2}$. The
second one results from the spontaneous nature of the $XX_{T_{\pm3}}^{0}$
radiative recombination and its re-excitation using CW light field. In order to estimate $T_{2}$, the second
contribution should be reduced to minimum. Using weak pulsed excitation
rather than CW, we previously showed that the coherence time of the
DE has a lower bound of about 100 ns\cite{Schwartz2015}. 

The obtained visibilities and DCP decay times  
are summarized in the upper and lower insets to Figure \ref{fig:7}(b), respectively. As expected,
the increase in the magnetic field does not affect the coherence of
the DE as clearly seen in the lower inset to Figure \ref{fig:7}(b). Clearly, the decay of the DCP ($T_{D}$) is almost field independent. 
Moreover, since the obtained $T_{D}$ of about $8$ns is about an order of magnitude shorter
than that measured under pulsed excitation in Ref {[}16{]}, one
can safely deduce that the dominant mechanism, which defines the
DCP oscillations decay time $T_{D}$ in our measurements is
the resonantly exciting laser field. 

In contrast, the upper inset to Figure \ref{fig:7}(b) shows that the visibility of the
DCP oscillations depends on the externally
applied field. This dependence is readily understood from Figure
\ref{fig:1}(b.ii) and Eqs. \ref{eq:9} and \ref{eq:10}, as resulting from   
the magnetic field induced changes of the eigenstates of the DE qubit. 
Symbols in the upper inset represent the measured visibility (derived from the first valley and second peak of the modeled DCP). 
The expected dependence from the numerical model is represented by a dash black line and that expected from Eq. \ref{eq:10} 
is represented by a solid orange line. The slight difference between the full numerical model and the simple analytical one 
is the absence of the effect of other levels 
(such as the DE biexciton) in the analytical model. 

In summary, we present an experimental and theoretical study of the
quantum dot confined dark exciton as a coherent two level system subject
to an externally applied magnetic field. Experimentally, we used polarization
sensitive intensity autocorrelation measurements of the optical transition
which connect a $XX_{{T_{3}}}^{0}$ biexciton state with the dark
exciton state. Detection of a circularly polarized photon from this
transition heralds the dark exciton and its spin state. By applying
an external magnetic field in the Faraday configuration, we measured
the Zeeman splitting of various lines and accounted for our measurements
by determining the g-factors of the electron, the hole and that of
two holes in a triplet spin state. We then used the external field
as a tuning knob for varying the dark exciton eigenstates. We showed
that this external control knob does not affect the long coherence
time of the dark exciton.
Theoretically, we were able to describe all our measurements using
a Lindblad type master equation model with a minimal number of free
fitting parameters. Ultimately, our work provides a better understanding
of the fundamentals of quantum dot excitations and may enable their
use in future technologies.

The support of the U.S. Israel Binational Science Foundation (BSF),
the Israeli Science Foundation (ISF), The Lady Davis Fellowship Trust
and the Israeli Nanotechnology Focal Technology Area on \textquotedblleft Nanophotonics
for Detection\textquotedblright{} are gratefully acknowledged. 

\bibliographystyle{unsrt}
\bibliography{DE_magnetic}

\end{document}